\documentclass[twocolumn,showpacs,preprintnumbers,amsmath,amssymb,superscriptaddress]{revtex4}
\usepackage{epsfig}
\usepackage{psfrag}
\usepackage{amsfonts}
\usepackage{graphicx}
\usepackage{dcolumn}
\usepackage{bm}

\begin{document}


\title{(Inverse) Magnetic Catalysis in Bose-Einstein Condensation of Neutral Bound Pairs }

\author{Bo  Feng}
\affiliation{
School of Physics, Huazhong University of Science and Technology, Wuhan 430074, China
}
\affiliation{
Institute of Particle Physics and Key Laboratory of Quark and Lepton Physics
(MOE),  Central China Normal University, Wuhan 430079, China}

\author{De-fu Hou}
\affiliation{
Institute of Particle Physics and Key Laboratory of Quark and Lepton Physics
(MOE),  Central China Normal University, Wuhan 430079, China}

\author{Hai-cang Ren}
\affiliation{
Physics Department, The Rockefeller University, 1230 York Avenue, New York, New York 10021-6399, USA}
\affiliation{
Institute of Particle Physics and Key Laboratory of Quark and Lepton Physics
(MOE),  Central China Normal University, Wuhan 430079, China}

\date{\today}

\begin{abstract}
The Bose-Einstein condensation of bound pairs made of oppositely charged fermions in a magnetic field is investigated.
We find that the condensation temperature shows the magnetic catalysis effect in weak coupling and the inverse magnetic
catalysis effect in strong coupling. The different
responses to the magnetic field can be attributed to the competition between the dimensional reduction by Landau orbitals in
pairing dynamics and the anisotropy of the kinetic spectrum of fluctuations (bound pairs in the normal phase).
\end{abstract}

\maketitle


\section{Introduction}

The behavior of a system consisting of charged fermions in a magnetic field had attracted considerable interests in recent years
especially in strongly interacting matter, where fundamental constituent-quarks exhibit a host of interesting
phenomena\cite{lectnotes}, such as Chiral Magnetic effect and Magnetic Catalysis of chiral symmetry breaking.
The latter one, which will be the main motivation to the present work, involves the dimensional reduction by the Landau orbitals
of charged fermions  under a magnetic field.
We shall investigate another (nonrelativistic) system  that shares the same physics, the Bose-Einstein Condensation (BEC) of composite bosons----neutral bound pairs made of two
oppositely charged fermions in the presence of an external magnetic field.

The underlying theory of strong interaction-Quantum Chromodynamics(QCD) possesses chiral symmetry for massless quarks, which is
spontaneously broken by a long range order because of the condensation of bound pairs formed by quark and antiquark. As the
density of states $D(E)\sim E^2$, with respect to the single quark energy $E$, vanishes at the Dirac point $E=0$ (analog of the
Fermi surface in a metal), a threshold coupling has to be attained for pairing. The terminology "Magnetic Catalysis" refers
to the fact that chiral symmetry is always spontaneously broken at finite magnetic field regardless of the coupling strength\cite{Klimenko, Miransky}. The physical reason of
this effect is the dimension reduction in the dynamics of fermion pairing in a magnetic field. The motion of charged
particle would be squeezed to a discrete set of Landau orbitals and is one-dimensional within each orbital. The system would thus become $1+1$ dimension when the magnetic field is sufficiently strong than the mass and energy of the fermions,  which would be restricted entirely in the lowest Landau level  (LLL) only.
Consequently, the density of states at the Dirac point becomes a nonzero constant proportional to the magnetic field $eB$. Such an enhancement would make the chiral condensate happen regardless of the interaction
strength, the magnetic field thus plays a role as the catalysis.  This is quite similar to the Bardeen-Cooper-Schrieffer (BCS) theory of superconductivity,
where a nonzero density of states at the Fermi surface supports Cooper pairing with an arbitrarily weak attraction.

It would be natural to expect a higher transition temperature from the chiral broken phase to the chiral symmetric phase due
to magnetic catalysis effect. This is indeed the case within mean-field approximations of effective model studies, it was found
that the chiral phase transition is
significantly delayed by a nonzero magnetic field even including the $\rho$ meson contribution\cite{Fukushima,Chernodub,Skoko}. The pseudo-critical temperature  of chiral restoration was also found to increase linearly with the magnetic field in a quark-meson model using the
functional renormalization group equation\cite{Kanazawab}. The recent lattice calculations\cite{latticetemp}, however, provide surprising results that the pseudo-critical
temperature of chiral restoration drops considerably for increasing magnetic field. On the other hand, the chiral condensate increase with
increasing magnetic field at low temperature consistent with magnetic catalysis while it turns out to be monotonously decreasing
at high temperature\cite{latticegap}, which is in apparent conflict with the magnetic catalysis and termed as inverse magnetic
catalysis evoking an extensive studies\cite{Schmitt,Fukushimapawlowski,Fukushimahidaka,Kojo,Bruckmann,Mei,Efrain,Pinto}.

While mean field approximation gives sensible results in certain circumstances, fluctuations can break it down, especially
in strong coupling domain or in lower dimensions. As was shown in \cite{CMWH} in the absence of magnetic field, a long
range order cannot survive at a nonzero temperature in the spatial dimensionality two or less because of the fluctuation of
its phase. A long wavelength component of the fluctuation variance goes like $1/p^2$ with $\bf p$ the momentum, which gives
rise to infrared divergence of the momentum integration in two and lower dimensions. The anisotropy introduced by a magnetic
field $B\bf{\hat z}$ renders the long wavelength fluctuation $\sim 1/(p_z^2+\kappa p_\perp^2)$, with $\kappa$ a positive constant
between zero and one. Such a distortion of the bosonic spectrum towards dimensionality one ($\kappa\to 0$), as a consequence
of the dimension reduction of the pairing fermions, would enhance the phase fluctuation. A preliminary study of the Ginzburg-Landau theory of the chiral
phase transition\cite{Ginzburgwindow} reveals the same effect and the Ginzburg critical window gets widened in the
presence of magnetic field, indicating the enhancement of the long wavelength fluctuations.

The BEC of bound pairs made of oppositely charged fermions in a magnetic field provides another
platform to explore the competition between the enhanced Cooper pairing by Landau orbitals and the enhanced phase fluctuation
by the distortion of the bosonic spectrum. Our system corresponds to the BEC limit
of the BCS/BEC crossover, which has been studied extensively in the absence of magnetic field for nonrelativistic fermions
\cite{Nozieres,Sademelo} and relativistic ones \cite{abuki,Dengjian,Efraincrossover,Zhuang}.
We follow the functional integral formulation developed in \cite{Sademelo} and calculate the leading (Gaussian) correction
to the effective action. A technical simplification in the BEC limit is that all summations over Landau orbitals involved can 
be carried out analytically, resulting in an explicit formula of the critical temperature under an aribitrary magnetic field.
We found that the critical temperature for the BEC was dramatically affected by the magnetic field exhibiting magnetic
catalysis or inverse magnetic catalysis depending on the coupling strength. In the weak
coupling domain, where no bound pairs(composite bosons) exist at zero magnetic field, the magnetic catalysis induces bound pairs and thereby
a BEC. The critical temperature increases with increasing magnetic
field. In the strong coupling domain, where bound pairs exist without magnetic field, an inverse magnetic catalysis was
found. The critical temperature decreases as increasing magnetic field, signaling the enhanced fluctuation in a magnetic
field.

The rest of the paper is organized as follows: in Section II we lay out the general formulation and present the mean field
approximation.  The fluctuations beyond the mean field theory, which is necessary for BEC, is calculated
under the Gaussain approximation in Section III. The magnetic field dependence of the BEC
temperature is investigated in Section IV. Section V is devoted to the conclusions and outlooks. Some calculation details and
useful formulas are presented in the Appendices A, B and C. Throughout the paper, we will work Euclidean signature with the
four vector represented by $x^\mu=(i\tau,{\bf x}), {q^\mu=(i\omega_n, {\bf q})}$ with $\omega_n$ the Matsubara frequency for
bosons $\omega_n=2i\pi nT$ and for fermions $\omega_n=(2n+1)i\pi T$.

\section{General Formulation and Mean Field Theory}

We consider a system consisting of nonrelativistic fermions of mass $m$ and chemical potential $\mu$ with opposite charge
interacting through a short ranged instantaneous attractive interaction. The Hamiltonian density reads
\begin{align}
\nonumber{\cal H}[\psi,\psi^\dagger]=&\sum_{\sigma=\pm}\psi_\sigma^\dagger(x)\left[\frac{\left(-i{\bf\nabla}+\sigma e{\bf A}\right)^2}{2m}-\mu\right]\psi_\sigma(x)\\
&-g\psi_+^\dagger(x)\psi_-^\dagger(x)\psi_-(x)\psi_+(x).
\label{hamiltonian}
\end{align}
where $e>0$ is the charge magnitude carried by each fermion, $\sigma=\pm$, $g>0$ and $\bf A$ is the vector potential underlying an external
magnetic field, $\bf B=\bf\nabla\times\bf A$.
To avoid the Meissner effect,  only fermions with opposite charges can pair. For the sake of simplicity, we ignore the
spin degrees of freedom. The thermodynamic potential density of the system reads
\begin{equation}
\Omega=-\frac{1}{\beta V}\ln{\cal Z}
\end{equation}
where $\beta=1/T$ and $V$ is the volume of the system.
The path integral representation of the partition function ${\cal Z}$ reads
\begin{equation}
{\cal Z}=\int {\cal D}\psi_\sigma^\dagger(x){\cal D}
\psi_\sigma(x)
\exp[{\cal S}].
\end{equation}
with the action ${\cal S}$ given by
\begin{equation}
{\cal S}=\int d\tau d^3{\bf x}\left(-\sum_{\sigma}\psi_\sigma^\dagger(x)
\frac{\partial}{\partial\tau}\psi_\sigma(x)-{\cal H}[\psi,\psi^\dagger]\right).
\end{equation}
where the Grassmann variables $\psi$ and $\psi^\dagger$ are antiperiodic in $\tau$ and {\it independent} of each other.
The number density of fermions is given by
\begin{equation}
n=-\left(\frac{\partial\Omega}{\partial\mu}\right)_{T,\bf B}.
\label{density1}
\end{equation}

Introducing the standard Hubbard-Stratonovich field $\Delta(x)$ coupled to $\psi^\dagger_+\psi^\dagger_-$, the partition function
is converted to
\begin{widetext}
\begin{align}
\nonumber{\cal Z}=&\int {\cal D}\psi_\sigma^\dagger(x){\cal D}\psi_\sigma(x){\cal D}\Delta^*(x){\cal D}\Delta(x)\exp\left\{\int d\tau d^3{\bf x}\left(-\psi_\sigma^\dagger(x)\frac{\partial}{\partial\tau}\psi_\sigma(x)-\psi_\sigma^\dagger(x)\frac{\left(-i\nabla+\sigma e{\bf A}\right)^2}{2m}\psi_\sigma(x)\right.\right.\\
&+\left.\left.\mu\psi_\sigma^\dagger(x)\psi_\sigma(x)+\Delta(x)\psi_+^\dagger(x)\psi_{-}^\dagger(x)+\Delta^*(x)\psi_-(x)\psi_+(x)-\frac{|\Delta(x)|^2}{g}\right)\right\}.
\end{align}
\end{widetext}
and becomes bilinear in fermion fields. In terms of the Nambu-Gorkov(NG) spinors
\begin{equation}
\Psi(x)=\left(\begin{array}{c}
\psi_+(x)\\
\psi^\dagger_-(x)\\
\end{array}\right),\hspace{0.2cm}
\Psi^\dagger(x)=\left(\psi^\dagger_+(x),
\psi_-(x)\right).
\end{equation}
the partition function becomes
\begin{align}
\nonumber{\cal Z}=&{\cal N}\int {\cal D}\Psi^\dagger(x){\cal D}\Psi(x){\cal D}\Delta^*(x){\cal D}\Delta(x)\exp\int d\tau d^3{\bf x}\\
&\left[\int d\tau^\prime d^3{\bf x}^\prime\Psi^\dagger(x)G^{-1}(x,x^\prime)\Psi(x^\prime)-\frac{\left|\Delta(x)\right|^2}{g}\right].\label{partitionfun}
\end{align}
with
\begin{align}
\nonumber G^{-1}=&\left[
\begin{array}{cc}
-\frac{\partial}{\partial\tau}-\frac{\left(-i\nabla+e{\bf A}\right)^2}{2m}+\mu & \Delta(x)\\
\Delta^*(x) & -\frac{\partial}{\partial\tau}+\frac{\left(-i\nabla+e{\bf A}\right)^2}{2m}-\mu\\
\end{array}\right]\\& \times\delta^4(x-x^\prime).\label{inversepropagator}
\end{align}
where ${\cal N}$ is a constant. Integrating out the fermionic NG fields, we obtain the parition function
\begin{equation}
{\cal Z}={\cal N}\int {\cal D}\Delta^*(x){\cal D}\Delta(x)\exp({\cal S}[\Delta(x)]),\label{pathintegral}
\end{equation}
with the action  $S$ given by
\begin{align}
{\cal S}[\Delta]=-\int d\tau d^3{\bf x}\frac{\left|\Delta(x)\right|^2}{g}+{\rm Tr}\ln G^{-1}(x,x^\prime),\label{partitionfunction}
\end{align}
where the trace in (\ref{partitionfunction}) is over space, imaginary time and NG indices.

For a uniform magnetic field $\bf B$ considered in this work, we choose the Landau gauge, in which the vector potential
is $A_x=A_z=0, A_y=Bx$ and the magnetic field is thus along $z$ direction and the system is translationally invariant. To explore the
long range order of the system, we make a Fourier expansion
\begin{equation}
\Delta(x) = \sqrt{\frac{1}{\beta V}}\sum_{\omega_{n_k},{\bf k}}
e^{-i\omega_{n_k}\tau+i{\bf k}\cdot{\bf x}}\Delta(i\omega_{n_k},{\bf k})= \Delta_0+\Delta'(x),
\label{fourier}
\end{equation}
where we have singled out the zero energy-momentum component of the expansion. Carrying out the path integral over
$\Delta'(x)$, we end up with
\begin{equation}
{\cal Z}={\cal N}\int {\cal D}\Delta_0^* {\cal D}\Delta_0\exp\left[-\beta V\Xi(|\Delta_0|)\right],
\end{equation}
and the thermodynamic potential density in the infinite volume limit equals to the value of the function
$\Xi(|\Delta_0|)$ at its saddle point $\bar\Delta_0$ determined by
\begin{equation}
\left(\frac{\partial\Xi}{\partial|\Delta_0|^2}\right)_{T,\mu,\bf B}=0.
\end{equation}
A nontrivial saddle point, $\bar\Delta_0\neq 0$, corresponds to a long range order and the superfluidity phase of
the system. $\bar\Delta_0$ drops to zero at the transition to the normal phase. Expanding the function $\Xi$
in a power series in $|\Delta_0|^2$,
\begin{equation}
\Xi(|\Delta_0|^2)=\Xi(0)+\alpha(T,\mu,{\bf B})|\Delta_0|^2+...,
\label{expand}
\end{equation}
where $\Xi(0)$, $\alpha(T,\mu,\bf B)$ and the coefficients of higher order terms of (\ref{expand}) include the
contribution from the fluctuation field
$\Delta'(x)$ defined in (\ref{fourier}). A negative value of the coefficient $\alpha(T,\mu.\bf B)$ signals the instability
of the normal phase, $\Delta_0=0$, and the critical temperature $T_c$, and the chemical potential $\bar\mu$ for
the instability satisfy the condition
\begin{equation}
\alpha(T_c,\bar\mu,\bf B)=0.
\label{gap1}
\end{equation}
The critical temperature at a given density is obtained by solving both eqs. (\ref{gap1}) and (\ref{density1}) simultaneously.

The mean field approximation ignores $\Delta'(x)$ and the eigenvalues of the inverse
propagator (\ref{inversepropagator}) with $\Delta(x)=\Delta_0$ can be easily found. We obtain
\begin{align}
\label{XiMF} \Xi(|\Delta_0|^2)=&\frac{1}{g}|\Delta_0|^2-\frac{1}{\beta V}\sum_n\sum_{k_y,k_z;l}\\
\nonumber&\ln
\left[(i\omega_n)^2-\left(\varepsilon_{k_z}+l\omega_B-\chi\right)^2-|\Delta_0|^2\right].
\end{align}
where $l=0,1,2,...$ are the Landau levels and $\varepsilon_{k_z}=k_z^2/2m$. We have also defined $\chi=\mu-\omega_B/2$ with
$\omega_B=eB/m$ the cyclotron frequency. The symbol $V^{-1}\sum_{k_y,k_z;l}$ is the abbreviation of
$eB/(2\pi)^2\sum_{l=0}^{\infty}\int_{-\infty}^{\infty}dk_z$. The coefficient $\alpha(T,\mu,\bf B)$ under the mean field
approximation can be readily extracted from the Taylor expansion of RHS of (\ref{XiMF}) and the condition (\ref{gap1})
becomes
\begin{equation}
\frac{1}{g}=\frac{1}{2V}\sum_{k_y,k_z;l}\frac{1}{\varepsilon_{k_z}+l\omega_B-\bar\chi}
\tanh\frac{\varepsilon_{k_z}+l\omega_B-\chi}{2T_c}.\label{gapeqwithT}
\end{equation}
In BCS limit, this equation would be solved to yield the critical temperature with the chemical potential given by that
of an ideal Fermi gas at a given density (the limit of eq.(\ref{density1}) with $\Omega=\Xi$ at $\Delta_0=0$, $T=0$ and ${\bf B}=0$).
In BEC limit, however, the role is reversed\cite{Sademelo}. Eq.(\ref{gapeqwithT}) determines the chemical potential. In the latter
case, the fluctuation contribution to $\Xi$ has to be restored to determine the critical temperature at a given density
through (\ref{density1}).

For negative $\chi$ with $T<<|\chi|$, the hyperbolic tangent function in (\ref{gapeqwithT}) may be approximated by one
and we end up with
\begin{equation}
-\frac{m}{4\pi a_s}=\frac{1}{2V}\left[\sum_{k_y,k_z;l}\frac{1}{\varepsilon_{k_z}+l\omega_B-\bar\chi}
-\sum_{\bf k}\frac{1}{2\varepsilon_{\bf k}}\right]
\label{tobereg}
\end{equation}
where we have introduced a renormalized coupling constant according to
\begin{equation}
\frac{1}{g_R}\equiv\frac{1}{g}-\frac{1}{V}\sum_{\bf k}\frac{1}{2\varepsilon_{\bf k}}\equiv -\frac{m}{4\pi a_s}
\label{renorm}
\end{equation}
with $a_s$ the $s$-wave scattering length extracted from the low energy limit of the two-body scattering in vacuum and
in the {\it absence} of a magnetic field so that the RHS is free from UV divergence. Carrying out the summation
explicitly (for details, see appendix A), we find that
\begin{equation}
-\frac{m}{4\pi a_s}=\frac{\sqrt{\omega_B}m^{3/2}}{4\sqrt{2}\pi}\zeta\left(\frac{1}{2},\frac{|\bar\chi|}{\omega_B}\right)
\label{gapeq},
\end{equation}
In obtaining this equation, the contributions from all Landau levels have been taken into account and this summation gives
rise to the Hurwitz zeta function, which was defined by
\begin{equation}
\zeta(s,a)=\sum_{n=0}^{\infty}\frac{1}{(n+a)^s}.
\end{equation}
for ${\rm Re}s>1$ and can be continuated to the entire $s$-plane with a pole at $s=1$ in terms of its integral representation.

The eq.(\ref{gapeq}) sets the chemical potential at the energy of a bound pair of zero center-of-mass momentum
in vacuum and this is the condition for the BEC of an ideal Bose gas. The contributions of the bound
pairs of nonzero momentum, however, is ignored here. Therefore the mean field approximation is not
sufficient and the contribution from the bound pairs with nonzero momenta to the density equation (\ref{density1}) has to be restored
to determine the transition temperature (the density will be set low enough to justify the approximation
$\tanh\frac{\varepsilon_{k_z}+l\omega_B-\bar\chi}{2T_c}\simeq 1$.).

In the absence of magnetic field, the RHS of (\ref{gapeq}) becomes $-m^{3/2}\sqrt{|\bar\chi|}/(2\sqrt{2}\pi)$ and
we have a solution $\bar\chi=-1/(2m a_s^2)$ only for $a_s>0$, which defines the strong coupling domain. The weak coupling
domain, $a_s<0$, however, entirely resides on the BCS side of the BCS/BEC crossover. When the magnetic field is turned on,
the RHS of (\ref{gapeq}) can take both signs and a solution emerges in the weak coupling domain. This is caused by the
dimensional reduction of the Landau orbitals, i.e., magnetic catalysis and the BEC limit can be approached in both
strong and weak coupling domains.

\section{Gaussian Fluctuation}

The Guassian approximation of the fluctuation effect maintains $\Delta'(x)$ to the quadratic order in the path integral (\ref{pathintegral}),
while including $\Delta_0$ to all orders.
To locate the pairing instability starting from the normal phase, where $\bar\Delta_0=0$, the Gauss approximation amounts to
replace ${\cal S}[\Delta]$ of (\ref{partitionfunction})
by its expansion to the quadratic order in the entire boson field $\Delta(x)$.
\begin{align}
\nonumber{\cal S}[\Delta]&\simeq{\cal S}_{\rm eff.}[\Delta]
={\cal S}[0]-\int d\tau d^3{\bf x}\frac{\left|\Delta(x)\right|^2}{g}\\
&-\int d\tau d\tau' d^3{\bf x}  d^3{\bf x'}
\left[G_+(x,x')\Delta(x')G_-(x',x)\Delta^*(x)\right]
\label{effectiveaction},
\end{align}
with
\begin{equation}
G_\pm(x,x')=\left[-\partial_\tau\mp\left(\frac{\left(-i\nabla+ e{\bf A}\right)^2}{2m}-\mu\right)\right]^{-1}\delta^4(x-x')
\label{propagator}.
\end{equation}
In terms of the Fourier transformation (\ref{fourier}),
\begin{equation}
{\cal S}_{\rm eff}[\Delta]={\cal S}[0]-\sum_{\omega_{n_p},{\bf p}}
\Gamma^{-1}(i\omega_{n_p},{\bf p})|\Delta(i\omega_{n_p},{\bf p})|^2\label{expansionofeffectiveaction}
\end{equation}
where the dependence of the coefficient $\Gamma^{-1}(i\omega_{n_p},{\bf p})$ on $T$, $\mu$ and $\bf B$ has been suppressed
and the thermodynamic potential density reads
\begin{equation}
\Omega=\Omega_0-\frac{1}{\beta V}\sum_{\omega_{n_p}, {\bf p}}\ln\Gamma(i\omega_{n_p}, {\bf p}).
\end{equation}
where $\Omega_0=-2/(\beta V)\sum_{k_y,k_z;l}\ln\left[1+\exp(\varepsilon_{k_z}+l\omega_B-\chi)\right]$ is the thermodynamic potential
of an ideal Fermi gas.
It follows that
\begin{equation}
\alpha(T,\mu,{\bf B})=\Gamma^{-1}(0,0),
\label{gap2}
\end{equation}
and
\begin{equation}
n=n_0+\frac{1}{\beta V}\frac{\partial}{\partial\mu}\sum_{\omega_{n_p}, {\bf p}}\ln\Gamma(i\omega_{n_p}, {\bf p}),
\end{equation}
with $n_0={2}{V}\sum_{k_y,k_z;l}\left[\exp(\beta(\varepsilon_{k_z}+l\omega_B-\chi)+1\right]^{-1}$ the fermionic
contribution to the density. Continuating $i\omega_{n_p}$ to an arbitrary real frequency $\omega$ according to the
prescription in
\cite{BaymMermin} and introducing a phase shift defined by
$\Gamma(\omega\pm i0,{\bf p})=|\Gamma(\omega,{\bf p})|\exp[\pm i\delta(\omega, {\bf p})]$,
the number equation can also be written as\cite{Sademelo}
\begin{equation}
n=n_0+\frac{1}{V}\sum_{\bf p}\int_{-\infty}^{\infty}\frac{d\omega}{\pi}n_B(\omega)
\frac{\partial\delta}{\partial\mu}(\omega,{\bf p}).
\label{numberequation}
\end{equation}
with $n_B(\omega)=(e^{\beta\omega}-1)^{-1}$ the Bose-Einstein distribution function.
The pair of equations, (\ref{gapeqwithT}) and (\ref{numberequation}), at zero magnetic field are widely employed in the
context of BCS/BEC crossover in the literature.

To calculate $\Gamma(\omega,{\bf q})$, we write $G_\pm(x,x')$ of (\ref{propagator}) in terms of the eigenvalues and
eigenfunctions of $G_\pm^{-1}$,
\begin{equation}
G_\pm(x,x^\prime)=\sum_{K}\frac{\psi_K(\tau,{\bf x})\psi^*_{K}(\tau^\prime,{\bf x}^\prime)}
{i\omega_{n_k}\mp\left(\varepsilon_{k_z}+l\omega_B-\chi\right)},
\end{equation}
with abbreviation $K=(\omega_{n};l,k_y,k_z)$ and the notation $\sum_K=(\beta V)^{-1}\sum_{\omega_{n_k}}\sum_{k_y,k_z;l}$.
The eigenfunction in Landau gauge reads
\begin{equation}
\psi_{K}(\tau,{\bf x})=\frac{1}{\sqrt{ L_yL_z}}e^{-i\omega_n\tau+i(k_yy+k_zz)}u_l\left(x-\frac{k_y}{eB}\right),
\end{equation}
where $L_y, L_z$ are the normalization lengths along $y$ and $z$ axes and the $u$-function is the wavefunction of a
harmonic oscillator given by
\begin{equation}
u_n(z)=\frac{(eB)^{\frac{1}{4}}}{\pi^{1/4}\sqrt{2^n\cdot n!}}e^{-\frac{eBz^2}{2}}H_n(\sqrt{eB} z)\label{ufunction}.
\end{equation}
with $H_n(z)$ the Hermite polynomial. The $u$-functions satisfy the orthonormality relation $\int
dxu_n(x)u_m(x)=\delta_{nm}$.


In terms of the Fourier components of $\Delta(x)$, the trace term in (\ref{effectiveaction}) becomes
\begin{widetext}
\begin{align}
\nonumber&{\rm tr}\left[G_+(x,y)\Delta(y)G_-(y,x)\Delta^*(x)\right]\\
\nonumber=&\sum_{K,l^\prime}\sum_{\omega_{n_p},{\bf p}}\sum_{p^\prime_x}\left[\int dx^\prime e^{ip_xx^\prime}u_l\left(x^\prime-\frac{k_y}{eB}\right)u_{l^\prime}\left(x^\prime-\frac{q_y}{eB}\right)\right]\left[\int dx e^{-ip^\prime_xx}u_l\left(x-\frac{k_y}{eB}\right)u_{l^\prime}\left(x-\frac{q_y}{eB}\right)\right]\\
&\times\frac{\Delta(i\omega_{n_p},{\bf p})}{i\omega_{n_k}-\left(\varepsilon_{k_z}+l\omega_B-\chi\right)}\frac{\Delta^*(i\omega_{n_{p}},p_x^\prime,p_y,p_z)}{i\omega_{n_q}+\left(\varepsilon_{q_z}+l^\prime\omega_B-\chi\right)}.
\end{align}
\end{widetext}
where $q=k+p$, $\omega_{n_q}=\omega_{n_k}+\omega_{n_p}$ and ${\bf p}=(p_x,p_y,p_z)$.
Upon shifting the integration $k_y$ to $k_y+eBx$, the last two $u$-functions will no longer be coordinate dependent
and the first two u-functions depend only on the relative
coordinates. The translational invariance becomes explicit then. It would be convenient to introduce the center of mass
coordinate $X=\frac{x^\prime+x}{2}$ and the relative one $r=x^\prime-x$ and we obtain that
\begin{widetext}
\begin{align}
\nonumber{\rm tr}\left[G_+(x,y)\Delta(y)G_-(y,x)\Delta^*(x)\right]
=&\sum_{\omega_{n_k},l,k_z;l^\prime}\sum_{\omega_{n_p},{\bf p}}\int dr e^{ip_xr}\left[
\int \frac{dk_y}{2\pi} u_l(r-\frac{k_y}{eB})u_{l^\prime}(r-\frac{q_y}{eB})u_l(-\frac{k_y}{eB})u_{-l^\prime}(\frac{q_y}{eB})\right]\\
&\times\frac{1}{i\omega_{n_k}-\left(\varepsilon_{k_z}+l\omega_B-\chi\right)}\frac{1}{i\omega_{n_q}+\left(\varepsilon_{q_z}+l^\prime\omega_B-\chi\right)}|\Delta(i\omega_{n_p},{\bf p})|^2.
\end{align}
\end{widetext}
Making the variable transformation $s=r-k_y/(eB)$ and $t=-k_y/(eB)$, we have
\begin{align}
\nonumber&{\rm tr}\left[G_+(x,y)\Delta(y)G_-(y,x)\Delta^*(x)\right]=\frac{eB}{2\pi}\sum_{\omega_{n_k},l,k_z;l^\prime}\sum_{\omega_{n_p},{\bf p}}\\
&\times\frac{|\Delta(i\omega_{n_p},{\textbf p})|^2|I_{ll'}(p_x,p_y)|^2}{[i\omega_{n_k}-\left(\varepsilon_{k_z}+l\omega_B-\chi\right)]
[i\omega_{n_q}+\left(\varepsilon_{q_z}+l^\prime\omega_B-\chi\right)]}\label{tracewithexponential},
\end{align}
with
\begin{equation}
I_{ll'}(p_x,p_y)=\int_{-\infty}^\infty d\xi e^{ip_x\xi}u_l(\xi)e^{-\frac{p_y}{eB}\frac{d}{d\xi}}u_{l'}(\xi),
\label{integral}
\end{equation}
where the identity
\begin{equation}
{\rm exp}\left(a\frac{d}{dx}\right)f(x)=f(x+a).
\end{equation}
with $f(x)$ an arbitrary function is employed. As is shown in Appendix B, the integral (\ref{integral}) can be
calculated explicitly with the aid of the
raising and lowering operators pertaining to the harmonic oscillator wave function $u_n(\xi)$,
\begin{align}
&\xi = \frac{1}{\sqrt{2eB}}(a+a^\dagger),\\
&\frac{d}{d\xi} = \sqrt{\frac{eB}{2}}(a-a^\dagger)
\label{aadagger}.
\end{align}
and we obtain that
\begin{equation}
|I_{ll'}|=\sqrt{\frac{l_<}{l_>}}e^{-\frac{p_\perp^2}{4eB}}\left(\frac{p_\perp^2}{2eB}\right)^{\frac{|l-l'|}{2}}
L_{l_<}^{|l-l'|}\left(\frac{p_\perp^2}{2eB}\right)
\label{laguerre}.
\end{equation}
where $l_<=\min(l,l')$, $l_>=\max(l,l')$, ${\bf p}_\perp=(p_x,p_y)$ and $L_n^\alpha(z)$ is the generalized Laguerre
polynomial. Combining (\ref{effectiveaction}), (\ref{tracewithexponential}) and (\ref{laguerre}) and carrying out the
summation over the Matsubara frequency $\omega_{n_k}$ in (\ref{tracewithexponential}), we end up with
\begin{widetext}
\begin{align}
\Gamma^{-1}(i\omega_{n_p},{\bf p})=\frac{eB}{2\pi}e^{-\frac{p_\perp^2}{2eB}}\sum_{l,l',k_z}
\left\{\frac{l_<}{l_>}\left(\frac{p_\perp^2}{2eB}\right)^{|l-l'|}\left[L_{l_<}^{|l-l'|}\left(\frac{p_\perp^2}{2eB}\right)\right]^2\right.
\left.\frac{n_F\left(\varepsilon_{k_z}+l\omega_B-\chi\right)-n_F\left(-\varepsilon_{q_z}-l'\omega_B+\chi\right)}
{-i\omega_{n_p}+(\varepsilon_{k_z}+\varepsilon_{q_z})+(l+l')\omega_B-2\chi}\right\}+\frac{1}{g}.
\label{inversegamma}
\end{align}
\end{widetext}
with $n_F(z)=(1+e^{\beta z})^{-1}$ the Fermi-Dirac distribution function. The isotropy perpendicular to
the magnetic becomes evident in $\Gamma^{-1}$. Setting $i\omega_{n_p}=0$ and ${\bf p}=0$, we verify the relation
$\Gamma^{-1}(0,0)=\alpha(T,\mu,\bf B)$ with $\alpha(T,\mu,\bf B)$ given by the mean field theory of the previous section
and vanishes at $T=T_c$ and $\mu=\bar\mu$ according to (\ref{gapeqwithT}).

For a negative $\chi$ with $\beta|\chi|>>1$, the case
considered in this work, the numerator on RHS of (\ref{inversegamma}) may be approximated by -1 and the integration
over $k_z$ can be carried out analytically. We have
\begin{align}
&\Gamma^{-1}(i\omega_{n_p},{\bf p})\\
\nonumber=&-\frac{m^{\frac{1}{2}}eB}{4\pi}e^{-\frac{p_\perp^2}{2eB}}\sum_{l,l'}
\frac{\frac{l_<}{l_>}\left(\frac{p_\perp^2}{2eB}\right)^{|l-l'|}\left[L_{l_<}^{|l-l'|}\left(\frac{p_\perp^2}{2eB}\right)\right]^2}
{\sqrt{\frac{p_z^2}{4m}-2\chi+(l+l')\omega_B+i\omega_{n_p}}}+\frac{1}{g}.
\label{gammafinal}
\end{align}

The singularity structure of $\Gamma$, with $i\omega_{n_p}$ continuated to the entire complext plane,
reflects the two-fermion spectrum. There will be an isolated real pole representing the two-body bound pair and
a branch cut along the real axis representing the continuum of two-fermion excitations. For sufficiently large $\beta|\chi|$,
the contribution to the density is dominated by the bound pair pole. We henceforth consider the expansion of (\ref{gammafinal})
around this pole, which is determined by $\omega=0, {\bf p}=0, \mu={\bar\mu}={\bar\chi}+\omega_B/2$ with ${\bar\chi}$
the solution to the mean field equation (\ref{gapeq}) and ${\bar\chi}<0, \beta|{\bar\chi}|\gg 1$, to the second order in terms of
$\bf p$ and first order in terms of $\omega, \mu-\bar\mu$. We obtain that
\begin{align}
\Gamma^{-1}\simeq a_1\left[-\omega-2(\mu-{\bar\mu})+\frac{p_z^2}{4m}\right]+a_2\frac{p_\perp^2}{4m}.
\end{align}
with
\begin{equation}
a_1=\frac{m^{3/2}}{16\pi\sqrt{2\omega_B}}\sum_{l=0}^\infty\left(l+\frac{|{\bar\chi}|}{\omega_B}\right)^{-\frac{3}{2}}
=\frac{m^{3/2}}{16\pi\sqrt{2\omega_B}}\zeta\left(\frac{3}{2},\frac{|{\bar\chi}|}{\omega_B}\right),
\label{eq_a1}
\end{equation}
and
\begin{align}
\nonumber a_2=&\frac{m^{3/2}}{\pi\sqrt{2\omega_B}}\sum_{l=0}^\infty\left(\frac{l+\frac{1}{2}}{\sqrt{l+\frac{|{\bar\chi}|}{\omega_B}}}-\frac{l}{2\sqrt{l-\frac{1}{2}+\frac{|{\bar\chi}|}{\omega_B}}}\right.\\
\nonumber&-\left.\frac{l+1}{2\sqrt{l+\frac{1}{2}+\frac{|{\bar\chi}|}{\omega_B}}}\right)\\
\nonumber=&\frac{m^{3/2}}{\pi\sqrt{2\omega_B}}\left\{\zeta\left(-\frac{1}{2},\frac{|{\bar\chi}|}{\omega_B}\right)-\zeta\left(-\frac{1}{2},\frac{1}{2}+\frac{|{\bar\chi}|}{\omega_B}\right)\right.\\
&+\left.\left(\frac{1}{2}-\frac{|{\bar\chi}|}{\omega_B}\right)\left[\zeta\left(\frac{1}{2},\frac{|{\bar\chi}|}{\omega_B}\right)-\zeta\left(\frac{1}{2},\frac{1}{2}+\frac{|{\bar\chi}|}{\omega_B}\right)\right]\right\}.
\label{eq_a2}
\end{align}
where the frequency $\omega$ is the continuation of the Matsubara frequency $i\omega_{n_p}$ to the neighborhood of
the pole. Obviously, the kinetic term becomes anisotropic with respect to the directions along and perpendicular to the
magnetic field because of the rotational symmetry breaking by the magnetic field.

The partition function (\ref{pathintegral}) under the Gaussian approximation of fluctuations may be written as
\begin{align}
{\cal Z}={\cal N}\int{\cal D}\phi^*{\cal D}\phi\exp\left\{\sum_{\omega_{n_p},{\bf p}}\phi_p^*\left(\omega-\omega_b+2\mu\right)\phi_p\right\}\label{partitionrescaled}.
\end{align}
where $\phi$ is the rescaled field of the fluctuation $\Delta$ and $\omega_b=-E_B+\omega_B+{p_z}^2/(4m)+\kappa p_\perp^2/(4m)$
is the bosonic dispersion relation with $E_B=-2{\bar\chi}$ the binding energy that is measured from the lowest Landau level.
We have also the explicit expression of the anisotropy factor
\begin{widetext}
\begin{equation}
\kappa\equiv a_2/a_1=16\frac{\zeta\left(-\frac{1}{2},\frac{|{\bar\chi}|}{\omega_B}\right)-\zeta\left(-\frac{1}{2},\frac{1}{2}+\frac{|{\bar\chi}|}{\omega_B}\right)
+\left(\frac{1}{2}-\frac{|{\bar\chi}|}{\omega_B}\right)\left[\zeta\left(\frac{1}{2},\frac{|{\bar\chi}|}{\omega_B}\right)-\zeta
\left(\frac{1}{2},\frac{1}{2}+\frac{|{\bar\chi}|}{\omega_B}\right)\right]}{\zeta\left(\frac{3}{2},\frac{|{\bar\chi}|}{\omega_B}\right)}.
\label{anisotropy}
\end{equation}
\end{widetext}
As is shown in the Appendix C, $\kappa \le 1$ for an arbitrary value of the ratio $|{\bar\chi}|/\omega_B$ and is a monotonically
increasing function of this ratio.

The partition function (\ref{partitionrescaled}) is nothing but an ideal Bose gas with anisotropy in kinetic term and
$\Gamma(\omega,{\bf p})$ is proportional to the boson propagator.
The condensation temperature is determined by setting the chemical potential in (\ref{numberequation}) at the solution of the
mean field equation (\ref{gapeq}), i.e. $\mu={\bar{\mu}}=\omega_B/2+\bar\chi$, and the phase shift there reads
\begin{equation}
\delta(\omega,{\bf p})=\pi\theta(\omega-\omega_b+2\bar\mu).
\end{equation}
where $\theta(x)$ is the Heaviside step function with $\theta(x\geq 0)=1$ and otherwise zero. It follows then that
\begin{equation}
n=2\int \frac{d^3\bf p}{(2\pi)^3}\left[\exp\left(\frac{p_z^2+\kappa p_\perp^2}{4mT_c}\right)-1\right]^{-1},
\end{equation}
where the $n_0$ term of eq.(\ref{numberequation}) is ignored with $T_c<<|\bar\chi|$. Consequently, the BEC
temperature is given by
\begin{equation}
T_c=\kappa^{\frac{2}{3}}T_c^0,\label{t ransitiontemperature}
\end{equation}
where
\begin{equation}
T_c^0=\left[\frac{n}{2\zeta(3/2)}\right]^{2/3}\frac{\pi}{m}.
\end{equation}
is the condensation temperature of an ideal Bose gas of the same density at zero magnetic field.

Beyond the Gaussian approximation, we have also calculated the quartic term, ${\cal S}_{\rm quartic}[\Delta]$, of the effective action (\ref{partitionfunction})
in the limit of low energy and momentum of $\Delta(x)$ and obtained a term
\begin{equation}
-\frac{3m^{\frac{3}{2}}\omega_B^{-\frac{3}{2}}}{64\sqrt{2}\pi}\zeta\left(\frac{5}{2},\frac{\bar\chi}{\omega_B}\right)
\sum_{\omega_{n_p},{\bf p}}|\Delta(i\omega_{n_p},{\bf{p}})|^4.
\end{equation}
to be added to eq.(\ref{expansionofeffectiveaction}). This term gives rise to a repulsive interaction between the bound pairs.

\section{Bose-Einstein Condensation in a Magnetic Field}

In this section, we shall explore the magnetic field dependence of the BEC temperature
(\ref{t ransitiontemperature}) for both strong coupling, $a_s>0$ and weak coupling, $a_s<0$.

As the mean-field equation
(\ref{gapeq}) and the formula (\ref{anisotropy}) depend on the ratio $r\equiv |{\bar\chi}|/\omega_B$ through the Hurwitz zeta
function, we shall begin with an examination of the two asymptotic behaviors $r\gg1$ and $r\ll1$ of the Hurwitz zeta function
$\zeta(s,r)$.

The large $r$ expansion follows from the Hermite formula
\begin{equation}
\zeta(s,r)=\frac{r^{-s}}{2}+\frac{r^{1-s}}{s-1}+2\int_0^\infty\frac{(r^2+y^2)^{-s/2}\sin s\theta}{e^{2\pi y}-1
}dy,
\end{equation}
with $\theta=\arctan(y/r)$, and reads
\begin{align}
\nonumber\zeta(s,r)\simeq&\frac{r^{-s+1}}{s-1}+\frac{r^{-s}}{2}+\frac{sr^{-s-1}}{12}-\frac{s(s+1)(s+2)r^{-s-3}}{720}\\
&+O(r^{-s-5}).
\end{align}
The negative value of $\zeta\left(1/2,r\right)$ in this limit leads us to the strong coupling domain via
the mean-field equation (\ref{gapeq})
\begin{equation}
\frac{1}{a_s}\simeq\sqrt{2m|{\bar\chi}|}-\frac{1}{2}\sqrt{\frac{m}{2|{\bar\chi}|}}\omega_B>0,
\end{equation}
If follows that the approximate binding energy
\begin{equation}
|{\bar\chi}|\simeq\frac{1}{2ma_s^2}(1+eBa_s^2),
\end{equation}
with $a_s<<\frac{1}{\sqrt{eB}}$. The anisotropy factor (\ref{anisotropy}) reads
\begin{equation}
\kappa\simeq 1-\frac{1}{16}\left(\frac{\omega_B}{|\bar\chi|}\right)^2\simeq 1-\frac{1}{4}(eB)^2a_s^4.
\end{equation}
and gives rise to a slight suppression of the condensation temperature according to (\ref{t ransitiontemperature}),
corresponding to an inverse magnetic catalysis.

The small $r$ behavior follows from the relation
\begin{equation}
\zeta(s,r)=r^{-s}+\zeta(s,1+r)\simeq r^{-s}+\zeta(s),
\end{equation}
which, for $s>0$, is dominated by the first term on RHS and corresponds to the lowest Landau level approximation in our
problem. The positivity of $\zeta\left(\frac{1}{2},\frac{|{\bar\chi}|}{\omega_B}\right)$ in this case, i.e. $|{\bar\chi}|<<\omega_B$,
together with the mean-field equation (\ref{gapeq}) implies a negative $a_s$ and thereby the weak coupling domain, i.e.
\begin{equation}
\frac{1}{a_s}\simeq-\sqrt{\frac{m}{2|{\bar\chi}|}}\omega_B<0.
\end{equation}
It follows that the binding energy
\begin{equation}
|{\bar\chi}|\simeq\frac{1}{2}m\omega_B^2a_s^2,\label{solution}
\end{equation}
is entirely induced by the magnetic field, as a consequence of the magnetic catalysis. In terms of the solution (\ref{solution}),
the inequality $|{\bar\chi}|<<\omega_B$ implies $|a_s|<<\frac{1}{\sqrt{eB}}$. The anisotropy factor
\begin{equation}
\kappa\simeq 8\frac{|{\bar\chi}|}{\omega_B}\simeq 4eBa_s^2<<1.
\end{equation}
in this case and maximizes the suppression of the condensation temperature.

Since $\zeta\left(1/2,r\right)$ is a monotonically decreasing function of $r$ and is negative (positive) for a large
(small) $r$, its zero, $r_c$, serves a demarcation between the strong coupling domain, where $a_s>0$ and
$|{\bar\chi}|/\omega_B>r_c$, and the weak coupling domain, where $a_s<0$ and $|{\bar\chi}|/\omega_B<r_c$. The value
of $r_c$ as well as the solution of the mean-field equation (\ref{gapeq}) and the condensation temperature for
$|{\bar\chi}|/\omega_B=O(1)$ can only be calculated numerically. We find $r_c\simeq0.303$,
\begin{equation}
|\bar\chi|\simeq r_c\omega_B\simeq 0.303\omega_B.
\label{largeB}
\end{equation}
and $\kappa\simeq 0.792$ as $B\to\infty$.

\begin{figure}
\includegraphics[height=7cm]{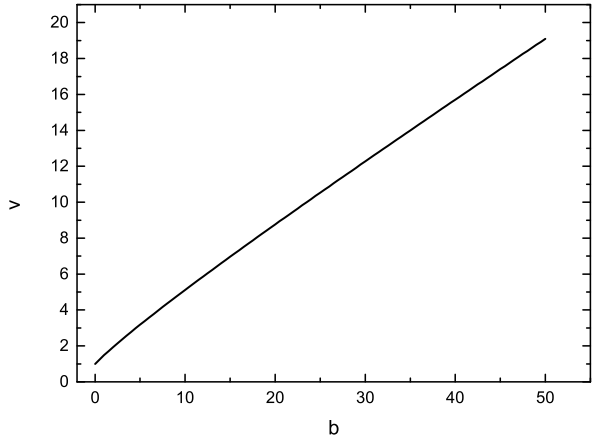}
\caption{\label{fig:epsart} The scaled binding energy $v$ versus the dimensionless magnetic field $b$ in strong coupling domain.}
\end{figure}

\begin{figure}
\includegraphics[height=7cm]{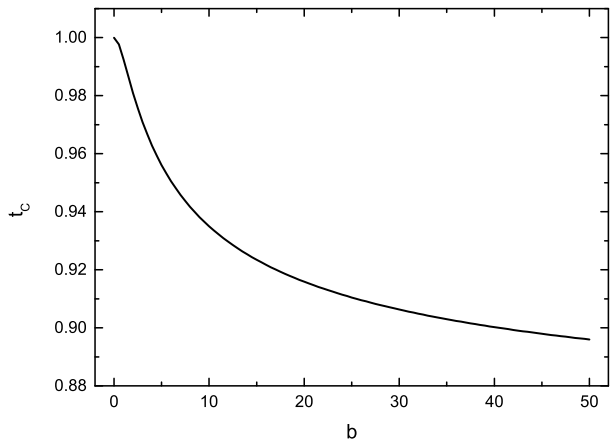}
\caption{\label{fig:epsart}The ratio of BEC temperature $t_c$ versus the dimensionless magnetic field $b$ in strong coupling domain.}
\end{figure}

In the strong coupling domain, $a_s>0$, bound pairs exist in the absence of magnetic field with the binding energy
$E_b=1/(ma_s^2)$ and condense at the temperature $T_c^0$. The mean-field equation (\ref{gapeq}) and the condensation temperature (\ref{t ransitiontemperature})
in a magnetic field can be expressed in terms of dimensionless quantities, i.e.
\begin{equation}
b^{-\frac{1}{2}}=-\frac{1}{2}\zeta\left(\frac{1}{2},\frac{v}{b}\right)
\label{gapeq1}
\end{equation}
and
\begin{equation}
t_c = \kappa^{\frac{2}{3}}\left(\frac{v}{b}\right),
\label{critical}
\end{equation}
where $b\equiv\frac{\omega_B}{E_b}$, $v\equiv\frac{|{\bar\chi}|}{E_b}$ and $t_c\equiv\frac{T_c}{T_c^0}$. The solution
of (\ref{gapeq1}) for $v$ and $t_c$ versus the dimensionless magnetic field $b$ are plotted in Fig.1 and Fig.2.
We find that the binding energy starts with a nonzero value at $b=0$, indicating the existence
of the bound pairs without magnetic field, and grows linearly for large $b$, consistent with the asymptotic behavior
(\ref{largeB}). The condensation temperature, however, deceases as magnetic field increases, consistent with the large $r$
limit. The physical reason for this inverse magnetic catalysis is the enhanced fluctuations by the anisotropic distortion of
the bosonic spectrum, $\kappa<1$, in the magnetic field. The effect is, however, rather mild with $\kappa$ decreasing
from one at $b=0$ to about 0.9 at $b=50$ because the ratio $r$ never drops to a level to warrant the LLL
approximation within the strong coupling domain.

In the weak coupling domain, $a_s<0$, bound pairs are formed through the mechanism of magnetic catalysis. The mean-field
equation becomes
\begin{equation}
b^{-\frac{1}{2}}=\frac{1}{2}\zeta\left(\frac{1}{2},\frac{v}{b}\right)
\label{gapeq2}
\end{equation}
with the sign on RHS opposite to that of (\ref{gapeq1}). The formula for the condensation temperature, (\ref{critical}),
remains unchanged. Here the denominator of $b$ and $t_c$, $|E_b|$ and $T_c^0$, do not carry direct physical meaning other than
reference scales because the bound pairs do not exist in the absence of magnetic field.
The solution of the mean-field equation for $v$ and $t_c$ versus $b$ in this case are plotted in Fig. 3 and Fig. 4. The
strong field limit of the binding energy also follows (\ref{largeB}). The difference, however, from the case in strong
coupling domain is that the binding energy at zero magentic field vanishes. The bound pairs exist only at nonzero magnetic
field, suggesting a BCS/BEC crossover induced by magnetic field. The condensation temperature in Fig.4 increases as magentic
field increases, which is consistent with the analysis in small $r$ limit. The LLL approximation works in the limit $r\to 0$,
where the anisotropy of the bosonic spectrum is maximized. An increasing magnetic field raises the ratio $r$ and promotes
the contribution from higher LL's, and thereby increases the condensation temperature.

\begin{figure}
\includegraphics[height=7cm]{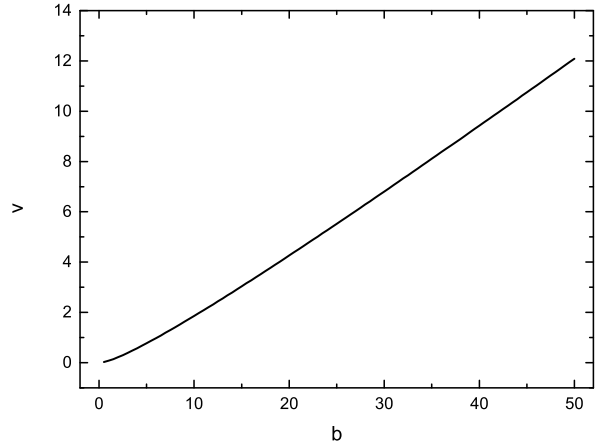}
\caption{\label{fig:epsart} The scaled binding energy $v$ versus the dimensionless magnetic field $b$ in weak coupling domain.}
\end{figure}

\begin{figure}
\includegraphics[height=7cm]{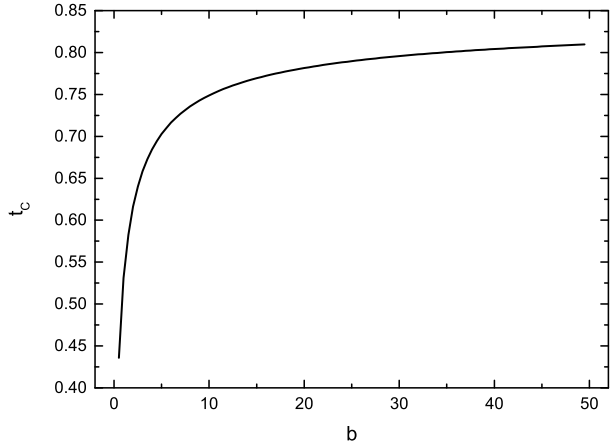}
\caption{\label{fig:epsart} The ratio of BEC temperature $t_c$ versus the dimensionless magnetic field $b$ in weak coupling domain.}
\end{figure}

Notice that, however, the condensation temperature is always suppressed compared with that of an ideal Bose gas of mass $2m$
regardless of the coupling strength because of the inequality $\kappa<1$ for all real $a_s$.

Before concluding this section, we would like to comment on the validity of the Gauss approximation of fluctuations in the
context of the BEC limit, which ignored the quartic and higher powers on $\Delta(x)$ in (\ref{partitionfunction}). These
terms represents the interactions among the Cooper pairs, which becomes significant when their wave functions overlap.
Therefore the approximation may deteriorate at the density at which the inter-particle distance $n^{-1/3}$
becomes comparable to the size of the bound pairs.

\section{Summary and Conclusions}

We have investigated a system of nonrelativistic bound pairs made of oppositely charged fermions in the presence of an external
magnetic field. We found that the variation of the BEC temperature with respect to the magnetic field depends on the
coupling strength of pairing. In strong coupling domain where the bound pairs(composite bosons) exist already without magnetic field, we found
the inverse magnetic catalysis that the condensation temperature decreases as increasing magnetic field. In weak coupling domain
where the bound pairs are induced by magnetic field, the transition temperature exhibits the usual magnetic catalysis effect.
In either domain, the condensation temperature is lower than that of an ideal Bose gas of the same mass, $2m$,
in the absence of
magnetic field. The suppression effect is maximized when the lowest Landau Level approximation works which requires the ratio
of binding energy relative to the lowest Landau
level over the spacing between adjacent Landau levels, $r=|\chi|/\omega_B<<1$. This condition is realized in
the weak coupling domain under a weak magnetic field. Otherwise, the ratio is order $O(1)$ and the suppression effect is less
pronounced. In particular, the binding energy diverges like $|\chi|\simeq 0.303\omega_B$ in the strong field limit, for both
strong and weak couplings, making the ratio 0.303 with the suppression factor $\kappa\simeq 0.792$. Of course, the BEC
approximation requires the fermion density of the system to be sufficiently low such that the bound pairs do not overlap.
With increasing density, individual bound pairs lose their identities and BCS condensation emerges. The fluctuations beyond
Gaussina approximation may also come to play then. This crossover
within the weak coupling domain under a magnetic field is what we called the magnetic field induced BCS/BEC crossover.
Without the magnetic field, the weak coupling domain corresponds only to the BCS side of the crossover.

To simplify the calculation, we ignored the spin degrees of freedom of the fermions as they do not contribute
to the pairing dynamics. The contribution of bound pairs with different spin configurations to the total density are
weighted by the Bose-Einstein distribution function with different Zeeman energies. For the temperature much lower than the
Zeeman energy of a fermion, the density is dominated by the pairing channel with the lowest Zeeman energy and our previous
results can be carried over.

It would be interesting to generalize our analysis to the relativistic fermions with the nonrelativistic propagators,
$G_\pm$ in (\ref{propagator}) replaced by Dirac propagators. The corresponding one loop diagram underlying $\Gamma^{-1}(\omega,\bf p)$
will be quadratically divergent. The leading divergence can be removed by the coupling constant renormalization like (\ref{renorm}), but
the logarithmic subleading divergence remains, which requires an explicit UV cutoff $\Lambda>>m$ of the pairing force. 
In addition to the weak coupling limit where the pairing dynamics is dominated by the lowest Landau level, the lowest Landau level
also dominates under an ultra strong magnetic field, $eB>>\Lambda^2$. The bosonic spectrum is expected to be highly 
anisotropic for $eB>>\Lambda^2$ with the critical temperature of BEC strongly suppressed by the fluctuations. The relativistic case is
current being investigated and the results will be reported elsewhere. The relativistic BCS/BEC crossover in a magnetic field was reported in \cite{Efraincrossover}
in the context of a boson-fermion model, where the boson is represented by an independent field with isotropic spectrum to the
zeroth order of coupling. The anisotropic distortion can only occur in higher orders there.

The Hamiltonian considered in the present work, eq. (\ref{hamiltonian}), is at the stage of a toy model and the conclusions 
are of theoretical values only. But the physics involved may be relevent to the color-flavor-locked phase or the single flavor 
planar phase of a dense quark matter in a strong magnetic field\cite{mcfl,planar}, where the pairing force stems from 
the non-perturbative QCD interaction.

\begin{acknowledgments}

The authors would like to thank I. Shovkovy, E. J. Ferrer and V. de la Incera for discussions and valuable comments. B. F. is supported by NSFC under grant No. 11305067
and the Fundamental Research Funds for the Central Universities, HUST: No. 2013QN015.
D. Hou and H.C. Ren are partly supported by NSFC under Grant Nos. 11375070, 11135011 and  11221504.
\end{acknowledgments}

\appendix

\section{The regularization in the mean field equation (\ref{tobereg})}

To regularize the RHS of the mean field equation (\ref{tobereg}), the summation over Landau orbitals is restricted to $l\le
N$ and, correspondingly, the transverse kinetic energy in the 2nd term (the sum over $\bf k$) is restricted below $N\omega_B$ i.e.
$\frac{1}{2}(k_x^2+k_y^2)\le N\omega_B$. The limit $N\to\infty$ will be taken in the end. Carrying out the momentum integral of
the 2nd term and the integration over $(k_y,k_z)$ of the 1st term, we find that
\begin{equation}
-\frac{m}{4\pi a_s} = \frac{m^{\frac{3}{2}}\sqrt{\omega_B}}{4\sqrt{2}\pi}\lim_{N\to\infty}
\left(\sum_{l=0}^N\frac{1}{\sqrt{l+\frac{\chi}{\omega_B}}}-2\sqrt{N}\right).
\end{equation}
To evaluate the limit, we introduce a sequence of analytic functions
\begin{equation}
f_N(s)\equiv \sum_{l=0}^N\left(l+\frac{|\chi|}{\omega_B}\right)^{-s}-\frac{N^{1-s}}{1-s},
\end{equation}
for positive integers $N$'s. The sequence converges uniformly in any closed domain with ${\rm Re}s>0$ and $s\neq 1$ and the
limit
\begin{equation}
f(s)=\lim_{N\to\infty}f_N(s),
\end{equation}
is therefore an analytic function within the same domain.
For ${\rm Re}s>1$, the limit of $N^{1-s}/(1-s)$ vanishes and we have
\begin{equation}
f(s)=\zeta\left(s,\frac{|\chi|}{\omega_B}\right).
\end{equation}
Following the principle of analytic continuation, we end up with
\begin{equation}
\lim_{N\to\infty}\left(\sum_{l=0}^N\frac{1}{\sqrt{l+\frac{\chi}{\omega_B}}}-2\sqrt{N}\right)
=f\left(\frac{1}{2}\right)=\zeta\left(\frac{1}{2},\frac{|\chi|}{\omega_B}\right).
\end{equation}
and eq.(\ref{gapeq}) follows.

\section{Calculation of the integral (\ref{integral})}

In this appendix, we show the details of the explicit calculation of the integral (\ref{integral}). In terms of the
raising and lowering operators (\ref{aadagger}), the integral can be written as
\begin{equation}
I_{ll'}=\langle l|e^{i\frac{p_x}{\sqrt{2eB}}(a+a^\dagger)}e^{-\frac{p_y}{\sqrt{2eB}}(a-a^\dagger)}|l'\rangle,
\end{equation}
with $u_l(\xi)=\langle\xi|l\rangle$. Using the operator relation
\begin{equation}
e^{A+B}=e^Ae^Be^{-\frac{1}{2}[A,B]},
\end{equation}
with $[A,B]$ commuting with both $A$ and $B$
twice, we find that
\begin{equation}
I_{ll'}=e^{i\frac{p_xp_y}{2eB}-\frac{1}{2}|w|^2}\langle l|e^{iw^*a^\dagger}e^{iwa}|l'\rangle,
\label{first}
\end{equation}
with
\begin{equation}
w=\frac{1}{\sqrt{2eB}}(p_x+ip_y).
\label{w}
\end{equation}
Expanding the exponential functions in $a$ and $a^\dagger$ and using the relation
$a|n\rangle=n|n-1\rangle$, we find
\begin{widetext}
\begin{align}
\nonumber \langle l|e^{iw^*a^\dagger}e^{iwa}|l'\rangle =& \sum_{n,n'\le l'}\frac{1}{n!n'!}\sqrt{\frac{l!l'!}{(l-n)!(l'-n')!}}i^{n+n'}
w^{*n}w^{n'}\langle l-n|l'-n'\rangle\\
\nonumber =& (iw^*)^{l-l'}\sum_{n'=0}^{l'}\frac{\sqrt{l!l'!}}{n'!(l-l'+n')!(l'-n')!}(-)^{n'}|w|^{2n'}\\
 =& \sqrt{\frac{l'!}{l!}}(iw^*)^{l-l'}L_{l'}^{l-l'}(|w|^2),
\label{case1}
\end{align}
\end{widetext}
for $l \ge l'$. For $l<l'$, we find
\begin{align}
\nonumber&\langle l|e^{i w^*a^\dagger}e^{iwa}|l'\rangle=\langle l'|e^{-iw^*a^\dagger}e^{-iwa}|l\rangle^*\\
&=\sqrt{\frac{l!}{l'!}}(iw)^{l'-l}L_l^{l'-l}(|w|^2).\label{case2}
\end{align}
Combining (\ref{first}), (\ref{w}), (\ref{case1}) and (\ref{case2}), we derive (\ref{laguerre}).

\section{The properties of the anisotropy factor $\kappa$}

To explore the properties of the anisotropy factor $\kappa$ as a function of $r=\frac{|\chi^*|}{\omega_B}$, we write
\begin{equation}
\kappa(r)=\frac{f(r)}{g(r)},
\end{equation}
with
\begin{align}
\nonumber f(r)=&16\left\{\zeta\left(-\frac{1}{2},r\right)-\zeta\left(-\frac{1}{2},\frac{1}{2}+r\right)\right.\\
&+\left.\left(\frac{1}{2}-r\right)\left[\zeta\left(\frac{1}{2},r\right)-\zeta
\left(\frac{1}{2},\frac{1}{2}+r\right)\right]\right\},
\end{align}
and
\begin{equation}
g(r)=\zeta\left(\frac{3}{2},r\right).
\end{equation}
It turns out that the Hermite formula is not convenient for this purpose and we start with the series representations
in (\ref{eq_a1}) and (\ref{eq_a2}). We have
\begin{widetext}
\begin{align}
\nonumber f(r) &= 16\sum_{l=0}^\infty\left(\frac{l+\frac{1}{2}}{\sqrt{l+r}}-\frac{l}{2\sqrt{l-\frac{1}{2}+r}}
-\frac{l+1}{2\sqrt{l+\frac{1}{2}+r}}\right)\\
\nonumber&= \frac{16}{\sqrt{\pi}}\sum_{l=0}^\infty\int_0^\infty dxx^{-\frac{1}{2}}\left[\left(l+\frac{1}{2}\right)e^{-(l+r)x}
-\frac{l}{2}e^{-\left(l-\frac{1}{2}+r\right)x}-\frac{l+1}{2}e^{-\left(l+\frac{1}{2}+r\right)x}\right]\\
 &= \frac{8}{\sqrt{\pi}}\int_0^\infty dxx^{-\frac{1}{2}}\frac{e^{-rx}}{\left(1+e^{-\frac{x}{2}}\right)^2}\ge 0,
\end{align}
\end{widetext}
where we have interchanged the order of integration and summation and have carried out the summation explicitly.
Likewisely
\begin{equation}
g(r)=\frac{2}{\sqrt{\pi}}\int_0^\infty dxx^{\frac{1}{2}}\frac{e^{-rx}}{1-e^{-x}}\ge 0.
\end{equation}
It follows that
\begin{equation}
f(r)-g(r)=\frac{8}{\sqrt{\pi}}\int_0^\infty dxx^{-\frac{1}{2}}\frac{e^{-rx}}{1-e^{-x}}
\left(\tanh\frac{x}{4}-\frac{x}{4}\right)\le 0,
\end{equation}
Therefore $f(r)\le g(r)$ and $\kappa(r)\le 1$. Taking the derivatives with respect to $r$, we find
\begin{equation}
\frac{df}{dr}-\frac{dg}{dr}=-\frac{8}{\sqrt{\pi}}\int_0^\infty dxx^{\frac{1}{2}}\frac{e^{-rx}}{1-e^{-x}}
\left(\tanh\frac{x}{4}-\frac{x}{4}\right)\ge 0,
\end{equation}
and then $\frac{df}{dr}\ge\frac{dg}{dr}$. Finally
\begin{equation}
\frac{d}{dr}\ln\kappa(r)=\frac{1}{f}\frac{df}{dr}-\frac{1}{g}\frac{dg}{dr}\ge 0.
\end{equation}
and $\frac{d\kappa}{dr}\ge 0$. The statements on $\kappa$ following (\ref{anisotropy}) are proved.

\newpage 


\end{document}